\documentclass{aa}

\usepackage{times}
\usepackage{psfig}
\usepackage{7766}

\def\3c{3C~273}
\def\z{\phantom{0}}
\def\BB{blue-bump}
\def\A{V_\nu}
\def\B{S_\nu}
\def\AAA{$\cal{B}$}
\def\BBB{$\cal{R}$}
\def\Bmin{{\cal R}^{\mathrm{min}}}
\def\Bmax{{\cal R}^{\mathrm{max}}}
\def\Fe2{Fe\,{\sc ii}}
\def\RS{R$_\mathrm{S}$}

\begin{document}
\thesaurus{3(11.17.4 3C 273; 13.21.1)}
\title{The \BB\ of \3c}
\author{S.Paltani\inst{1,2} \and T.J.-L.Courvoisier\inst{1,2} \and R.Walter\inst{2}}
\institute{Geneva Observatory, ch. des Maillettes 51, CH-1290 Sauverny,
Switzerland
\and
INTEGRAL Science Data Centre, ch.\ d'Ecogia 16, CH-1290 Versoix, Switzerland}
\offprints{St\'ephane Paltani (ISDC)}
\mail{Stephane.Paltani@obs.unige.ch}
\date{Received 11 May 1998 / Accepted 9 September 1998}
\maketitle

\begin{abstract}
  We present optical and ultraviolet observations of \3c\ covering the
  whole life of the IUE satellite. We analyze the variability
  properties of the light curves, and find that two variable
  components, written \AAA\ and \BBB\ respectively, must contribute to
  the \BB\ emission in this object.
  
  The \AAA\ component produces most of the variability in the
  ultraviolet domain. A maximum time scale of variability of about 2
  yr identical at all wavelengths is found. If discrete events produce
  this component, the event rate is $\sim$3-30 yr$^{-1}$.  Assuming an
  isotropic emission, each event must liberate about $10^{51}$ erg in
  the form of optical-to-ultraviolet radiation.  The spectral
  properties of the \AAA\ component suggest that reprocessing on a
  truncated disk, or partially-thick bremsstrahlung may be the
  emission mechanism. We find evidence of a lag of a few days between
  the light curves of the \AAA\ component at optical and ultraviolet
  wavelengths. An irradiated geometrically-thick accretion disk model
  satisfies all the constraints presented here.
  
  Neither the variability properties, nor the spectral properties of
  the \BBB\ component can be accurately measured. This component
  varies on very long time scales, and may completely dominate the
  historical light curve of \3c\ in the optical domain. Combining
  knowledge from other wavelengths, we obtain several indications that
  this component could reveal the ``blazar'' \3c. In particular, the
  light curve of this component is very similar to the hard X-ray
  light curve.
  

\end{abstract}

\section{Introduction}
\label{sec:intro}
Several components contribute to the multiwavelength continuum
emission of active galactic nuclei (AGN). One of these components,
which is called the ``\BB'', contains the dominant fraction of the
total emitted power in Seyfert 1 galaxies and (non-blazar) quasars.
Understanding the emission process at its origin would therefore be a
fundamental step towards the comprehension of the AGN phenomenon.

It is very easy to observe the \BB\ of the brightest Seyfert galaxies
and quasars with optical telescopes. One of the first result of these
observations is that these objects vary following complicated patterns
\cite[e.g.,][]{Press@78@flicker}. This has been the motivation for
important monitoring campaigns of the \BB\ emission of a few bright
sources, among which the quasar \3c. The aim of these campaigns was to
constrain the physics of the \BB\ using variability properties, either
of the lines, or of the continuum. In this paper, we concentrate on
the continuum emission, neglecting the emission line
spectrum.

Very detailed models for the continuum emission of the \BB\ have been
proposed, which can reproduce very correctly the
optical-to-ultraviolet spectral energy distribution (and sometimes up
to the soft X-ray, or even the hard X-ray, domain). They are often
based on the presence of an accretion disk
\cite[]{ShakuraSunyaev@73@ad}, and include the effect of
Comptonization \cite[]{CzernyElvis@87@discompt,RossAl@92@viscdisk},
the illumination by an X-ray source
\cite[]{Collin@91@reproc,RossFabian@93@xreproc}, or a corona
\cite[]{CzernyElvis@87@discompt,HaardtMaraschi@91@2phase,HaardtAl@94@blob}.
More exotic models invoke optically thin emission
\cite[]{Barvainis@93@free}, or supernova explosions
\cite[]{Terlevich@92@agnsn}.  The reproduction of a particular
spectrum is only the first requirement that a model has to fulfill. A
further requirement is that the model can explain a series of
simultaneous spectra. Such attempts to check whether the
above-mentioned models satisfy to this second requirement have been
made by \cite{RokakiAl@92@fitrep,Rokaki@93@55lab}, \cite{PW@96@pwrlaw}
(hereafter PW96) and \cite{LoskaCzerny@97@bbvar5548}.  The final step
would be to make sure that the complete dynamics of the spectra can be
reproduced by the models, including variability time scales, spectral
variations, and possible delays between \BB\ light curves.

In this paper we examine in detail the variability properties of the
continuum emission of \3c. We have gathered nearly 20 years of
observations with the \emph{International Ultraviolet Explorer} (IUE)
satellite and 10 years with an optical ground-based telescope. We
perform temporal analyses to understand how \3c\ varies, and we use
the temporal and spectral knowledge gained in these analyses to
constrain the physics of the emission at these wavelengths. Our
approach differs from those used in the papers cited above (although
we shall extend the work done in PW96) in that we focus on the
time-series properties of the \BB\ light curves, without neglecting
the spectral properties.  Our goal is to achieve here one of the most
complete analysis of the optical and ultraviolet emission ever done on
an active galactic nucleus, and to express the most stringent possible
constraints that are hidden in the huge amount of data collected on
\3c.

\section{The \BB\ continuum light curves}
\label{sec:bblight}

The data used in this paper are the IUE and optical observations
(Geneva photometry only) described in \cite{TPCAl@98@3cdata}. Details
about the data are to be found in this paper. For the IUE spectra, we
used the IUESIPS data reduction, as the Final Archive was far from
completion when we started the present analysis (Note however that the
results are very similar with the new reduction). The IUE spectra
identified as dubious in \cite{TPCAl@98@3cdata} have been discarded .
We corrected all the light curves for the effect of the reddening.  We
used the reddening law of \cite{SavageMathis@79@pouss}.  The value of
$E_{\mathrm{B-V}}$ has been set to 0.038, which is the value obtained
in PW96. We finally obtained 191 IUE SWP spectra, 155 IUE LWP/LWR
spectra, and 376 Geneva photometry optical observations.

We define 22 windows in the IUE spectra (7 for the SWP wavelength
range and 15 in the LWP/LWR wavelength range) of width 50 \AA, or 100
\AA\ (at the beginning and the end of the LWP/LWR wavelength range,
because of the large noise in these regions), carefully avoiding
emission lines.  It is unfortunately not possible to obtain optical
light curves free of emission line contamination, as they have been
obtained with broad-band photometry. However the contribution of the
emission lines has been estimated to be at most of the order of 10 \%
\cite[]{CAl@90@3c88}.  As Ly\,$\alpha$ variability is smaller than 10
\% \cite[]{UlrichAl@93@lyvar3c}, we do not expect that the optical
light curves are significantly perturbed by the contamination. The
average fluxes in these bins are plotted on Fig.~\ref{fig:sigma}.

\begin{figure}[p]
  \hspace*{-10mm}\mbox{\psfig{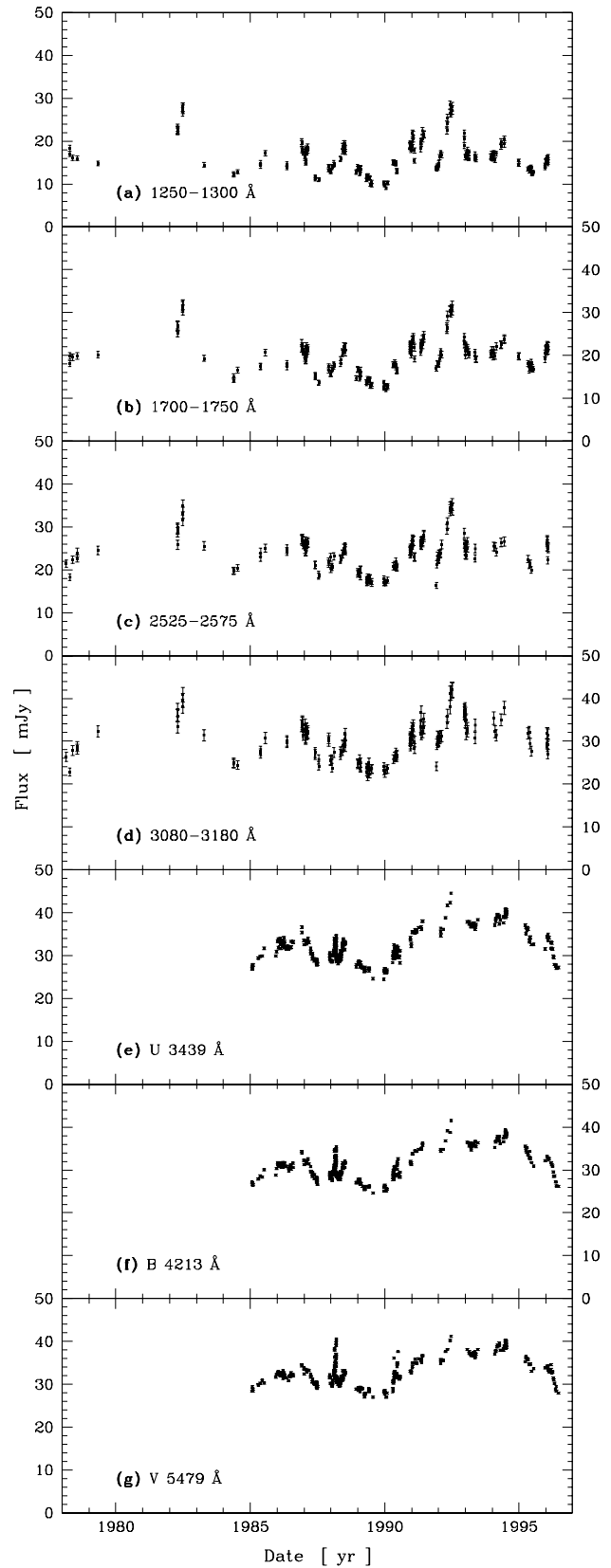}}
  \caption{Ultraviolet and optical light curves from our database
    dereddened with $E_{\mathrm{B-V}}=0.038$.  {\bf (a)--(d)} IUE
    light curves.  {\bf (e)--(g)} Optical light curves from the Swiss
    telescope.}
  \label{fig:lightcurves}
\end{figure}

Fig.~\ref{fig:lightcurves}a--d shows 4 of the ultraviolet light curves
extracted from the IUE spectra.  Fig.~\ref{fig:lightcurves}e--g shows
three optical light curves, in the U, B and V filters respectively.
One can see that all the ultraviolet and optical light curves are
similar. It is however apparent to the eye that a large bump is
present in the optical light curves (and in a less clear way in the
3\,080--3\,180 \AA\ light curve), which is (almost) absent at
1\,250--1\,300 \AA.

\section{Rapid flares in 1988 and 1990}
\label{sec:rapid}
\begin{figure}[tb]
  \mbox{\psfig{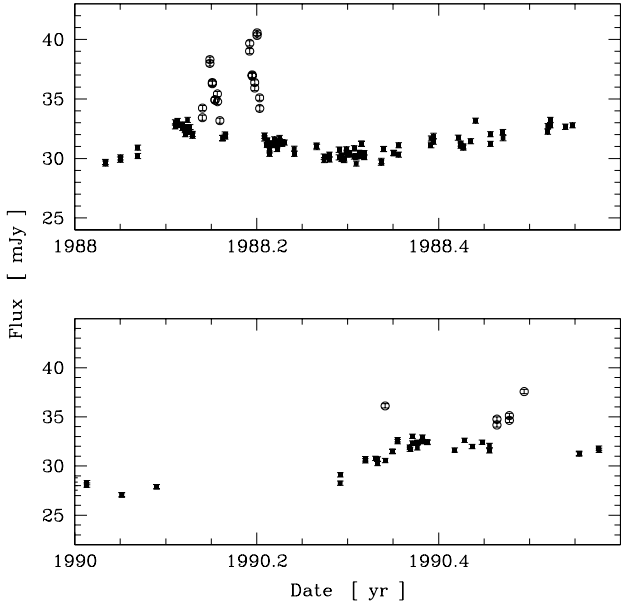}}
  \caption{Rapid flares in the V light curve of \3c. {\bf (a)} in
    Spring 1988.  {\bf (b)} in Summer 1990 (see text). The circles
    represent data in which the contribution of the flares is
    important. These data have been discarded from the rest of the
    present analysis}
  \label{fig:rapid}
\end{figure}

In 1988, \cite{CAl@88@vflare} noted the occurence of several very
rapid flares in the optical and near-infrared emission of \3c.
Comparable, but somewhat weaker, events have occurred between May and
June 1990. The details of the optical light curve are shown in
Fig.~\ref{fig:rapid}.  Unfortunately, the 1990 flares have been
detected several weeks after the observations, and no special
observation campaign has been performed.

\begin{figure}[tb]
  \hspace*{2.5mm}\mbox{\psfig{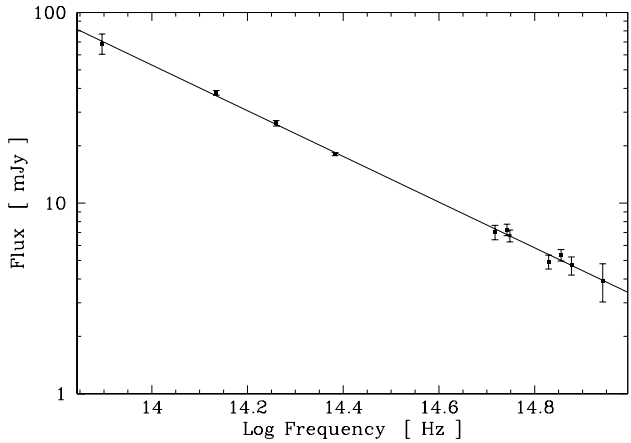}}
  \caption{Optical and near-infrared spectrum of the flare of March
    11, 1988.  The least-square linear regression has a slope $-1.2$
    (solid line).}
  \label{fig:rapspec}
\end{figure}

On March 11, 1988, \3c\ has been observed almost simultaneously
(within a couple of hours) in the Geneva photometry and in
near-infrared J, H, K, and L' photometry \cite[]{CAl@88@vflare}. If
one substracts the slowly varying continuum, one gets a power-law
spectrum with a spectral index of 1.2 ($f_\nu\sim \nu^{-\alpha}$), as
shown in Fig.~\ref{fig:rapspec}, which is much redder than the mean
optical spectral index, which is about 0.4. These flares have been
discussed at length elsewhere
\cite[]{CAl@88@vflare,RobsonAl@93@imr3c}, and will not be discussed
further here. The observations marked by open circles in
Fig.~\ref{fig:rapid} present colour indices that depart significantly
from the values obtained in adjacent observations, and hence are
supposed to be obtained when \3c\ was in a flare state. They are
discarded in the rest of the present analysis.

\section{Variability properties of the \BB}
\label{sec:varbb}

\subsection{Statistics of the light curves}
\label{sec:varampl}
\begin{figure}[tb]
  \mbox{\psfig{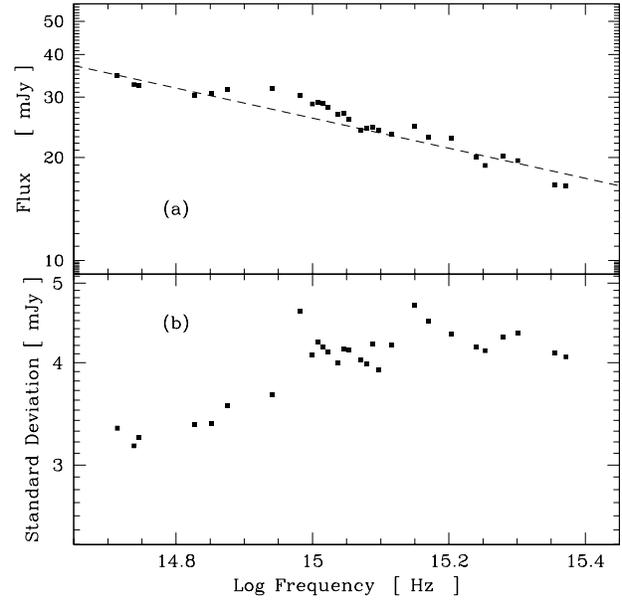}}
  \caption{{\bf (a)} Mean ultraviolet-to-optical continuum spectral
    energy distribution of \3c. The dashed line has a slope $-0.44$.
    {\bf (b)} Standard deviation of the measurements in each band}
  \label{fig:sigma}
\end{figure}

Fig.~\ref{fig:sigma}a shows the mean ultraviolet-to-optical spectrum
of \3c\ as a function of the wavelength. It presents a small bump
around $10^{14.9}$ Hz, which may be attributed to \Fe2\ and Balmer
line and continuum emission.  The rest of the spectrum is not very
different from an unique power-law with an index $0.44$.

On Fig.~\ref{fig:sigma}b, we can see the variability spectrum, i.e.\ 
the standard deviation as a function of the frequency. Contrarily to
what is seen in the mean spectrum, an extrapolation of the ultraviolet
variability clearly overestimates the variability in the V band. We do
not believe that this feature can have an instrumental origin, as the
transition between the ultraviolet and optical observations is smooth;
this cannot be the result of a discrepancy in the absolute calibration
between the two instruments, as it would produce a discontinuity.  The
important noise at the edges of the LWP/LWR wavelength range can be
easily seen in Fig.~\ref{fig:sigma}.

\subsection{Correlation analysis}
\label{sec:corr}
We compare the light curves with each other using correlation
analysis. In the case of astronomical discrete time series, two
methods are generally used: the discrete correlation function
\cite[DCF]{EdelsonKrolik@88@dcf} and the interpolated
cross-correlation function \cite[ICCF]{GaskellPeterson@87@gscc}. We
use here the ICCF method, because it gives better results in case of
the dense sampling available here
\cite[]{WhitePeterson@94@commcc,LitchfieldAl@95@simcc}.  We have
however checked that the results obtained with both methods are
qualitatively identical. 

\begin{figure}[tb]
  \mbox{\psfig{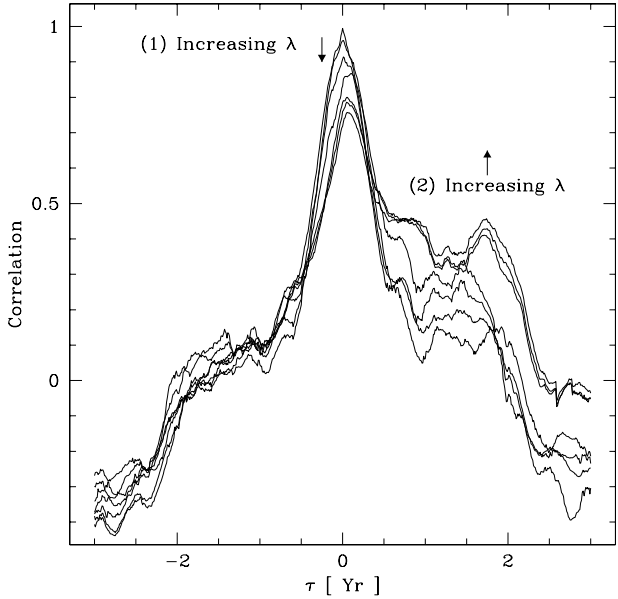}}
  \caption{Correlations between the IUE 1\,250-1\,300 \AA\ light curve
    and other IUE and optical light curves (successive solid lines).
    The effect of increasing the wavelength of the second light curve
    is shown for the peak at 0 lag (1), and for the hump (2)}
  \label{fig:correl}
\end{figure}

Fig.~\ref{fig:correl} shows the correlations between the 1\,250-1\,300
\AA\ light curve and the seven light curves from
Fig.~\ref{fig:lightcurves} as a function of the lag $\tau$.  Two
features emerge from these correlations. First, a peak close to
$\tau=0$. Its amplitude decreases with the wavelength, but is still
large ($\sim$ 0.75) in the correlation with the optical V light curve.
The width of the peak does not show any evidence of broadening when
the wavelength increases. It means that an important fraction of the
emission in the long-wavelength light curves varies in unison with the
shortest-wavelength light curve. One has however the impression that
the central peak is shifted towards larger values of $\tau$, and the
more so as the wavelength of the second light curve increases.  This
is an indication of a possible (short) lag between the light curves.
This is analyzed in more details in Sect.~\ref{sec:lag}.

The second feature is a hump between 0 and 2 yr.  Correlation analysis
must be taken with caution when dealing with discrete,
unevenly-sampled time series, as features without any physical reality
can be generated by sampling effects.  We have first attributed this
second feature to this phenomenon.  However, its amplitude increases
monotonically with the wavelength, a behaviour that cannot be due to
the sampling.  We shall discuss further this feature in
Sect.~\ref{sec:aabb}.

\subsection{Variability time scales}
\label{sec:tscal}
\begin{figure}[p]
  \hspace*{-10mm}\mbox{\psfig{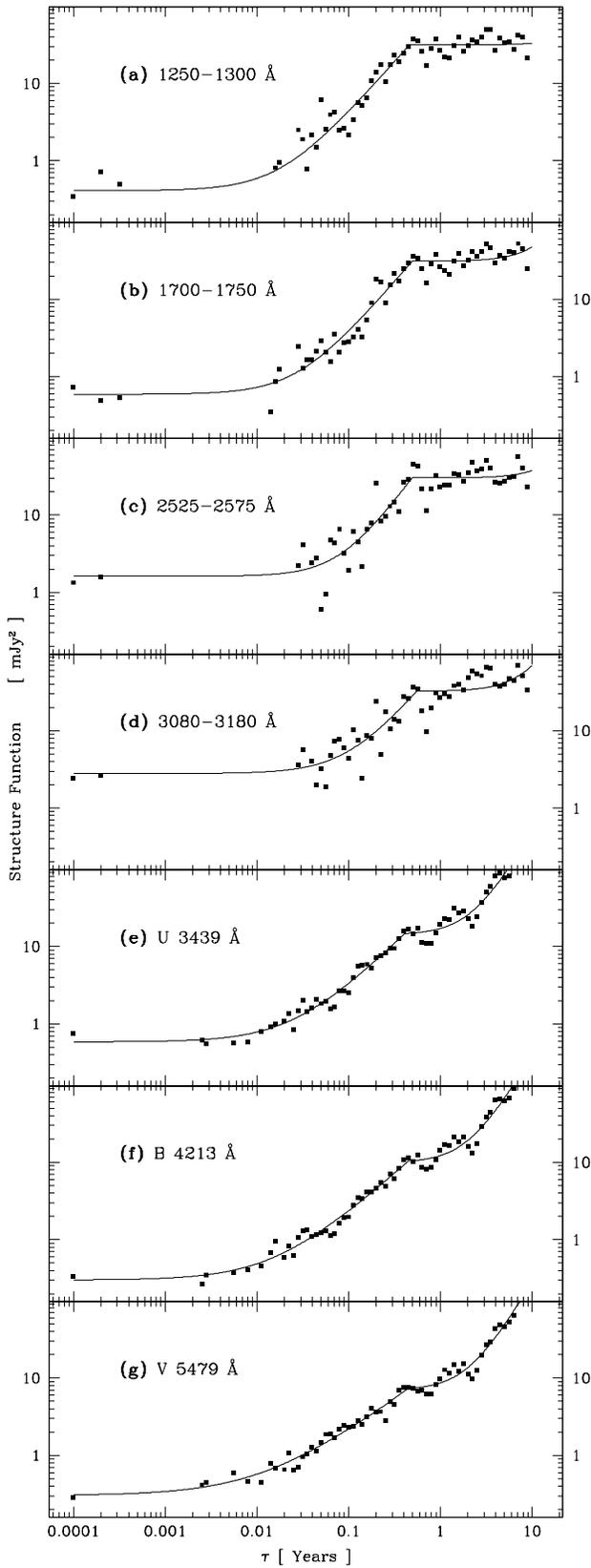}}
  \caption{{\bf a--g} Structure functions of 7 ultraviolet and optical
    light curves. The solid line is the best fit with the function
    given by Eq.~(\protect\ref{equ:funcsf}).}
  \label{fig:sf7}
\end{figure}

We investigate the variability time scales of the \BB\ of \3c\ using
structure function analysis. The structure function approach has been
introduced in astronomy by \cite{SimonettiAl@85@sfastro}. The
first-order structure function of a light curve (referred to as ``SF''
hereafter) is formally equivalent to its autocovariance. It however
highlights some important properties of the light curve in a much
clearer way. The SF of a time series $x(t)$ is a function of a time
lag ``$\tau$'', and is defined by:
\begin{equation}
  \label{eq-sf}
  \mathrm{SF}_{x(t)}(\tau)=\left<\left(x(t+\tau)-x(t)\right)^2\right>
\end{equation}
where $<\!z(t,\ldots)\!>$ is the average of $z$ over $t$.  We have
calculated SFs for the 29 optical and ultraviolet continuum light
curves.  The result is presented on Fig.~\ref{fig:sf7} for seven of them.

Let us assume that a time series $x(t)$ has a non-zero Fourier
power-spectrum only at frequencies larger than $f_{\mathrm{min}}$.
Below $\tau_{\mathrm{max}}= (f_{\mathrm{min}})^{-1}$, the larger
$\tau$, the larger the part of the power-spectrum that can make the
time series vary; thus the SF increases with $\tau$. Above
$\tau_{\mathrm{max}}$, all the power of the Fourier spectrum can be
used to make $x(t)$ vary, and the SF remains constant at a value equal
to twice the variance of $x(t)$.  One can convince oneself of this
result by noting that $x(t)$ and $x(t+\tau_0)$ with
$\tau_0>\tau_{\mathrm{max}}$ are completely uncorrelated, and
therefore their variances add up.  Similarly, if one observes
$y(t)=x(t)+\varepsilon(t)$, where $\varepsilon(t)$ is a white noise
representing the measurement uncertainties, then a constant is added
to the SF equal to twice the variance of $\varepsilon(t)$.  This
constant comes from the fact that $\varepsilon(t)$ and
$\varepsilon(t+\tau_0)$ are completely uncorrelated, $\forall
\tau_0\neq 0$.

In Fig.~\ref{fig:sf7}a, we observe that a plateau is reached around
$\tau\simeq 0.5$ years. However, in the SFs at longer wavelengths
(especially in the optical domain; cf.\ Figs~\ref{fig:sf7}d--f), the
SFs still increase for $\tau\simeq 10$ years. We have checked that
this was not due to the difference in time spans between the
ultraviolet and optical light curves by comparing the SF of a
truncated ultraviolet light curve with that of the total light curve.
Both SFs are indeed undistinguishable. To extract the quantitative
information contained in the SFs, one may wish to fit them with some
analytical function, and, if possible, the same for all the light
curves. We have found one function that fits rather well all SFs, and
whose parameters may have a physical interpretation.  This function is
given by:
\begin{equation}
  \label{equ:funcsf}
  f(\tau)= 2\cdot p_0^2 + \left\{
     \begin{array}{ll}
       p_2\cdot(\frac{\tau}{p_1})^{p_3},&\mathrm{if~} \tau<p_1\\
       p_2,&\mathrm{if~} \tau\ge p_1\\
     \end{array}
   \right\} + p_4\cdot\tau^2
\end{equation}
This function contains three separate terms:
\begin{itemize}
\item the measurement uncertainty, given by $p_0$, whose contribution
  to the SF is twice its square.
\item the \AAA\ term (in anticipation to Sect.~\ref{sec:twocomp}),
  whose SF is a power-law that increases with an index $p_3$ until a
  plateau is found at $\tau=p_1$. After the plateau, this term remains
  constant at a value $p_2$.
\item The \BBB\ term (again, see Sect.~\ref{sec:twocomp}), whose SF is
  a power-law with an index 2, which is the maximum index that can be
  exhibited by a SF. It describes a long-term variation.
\end{itemize}

\begin{figure}[t]
  \vspace*{10pt}
  \mbox{\psfig{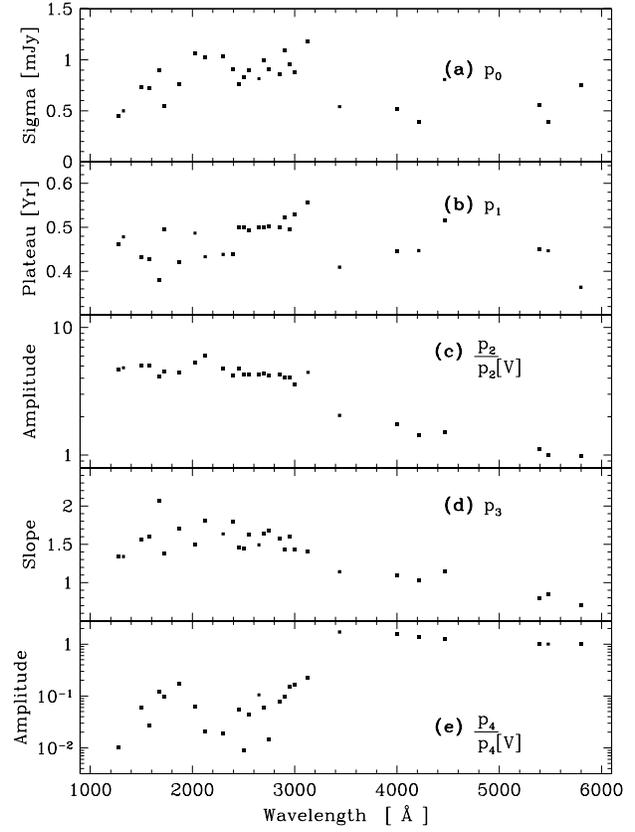}}
  \caption{Parameters derived from the SFs. {\bf (a)} $p_0$, the
    measurement uncertainty. {\bf (b)} $p_1$, the location of the
    plateau. {\bf (c)} $p_2/p_2[V]$, the amplitude of the \AAA\ term
    relatively to its amplitude in the V filter. {\bf (d)} $p_3$, the
    slope of the increasing part of the \AAA\ term.  {\bf (e)}
    $p_4/p_4[V]$, the amplitude of the \BBB\ term relatively to the
    its amplitude in the V filter.}
  \label{fig:par}
\end{figure}

The justification of Eq.~(\ref{equ:funcsf}) is that the SF is linear,
i.e.\ $\mathrm{SF}_{x(t)+y(t)}(\tau)= \mathrm{SF}_{x(t)}(\tau)+
\mathrm{SF}_{y(t)}(\tau)$, if $x(t)$ and $y(t)$ are completely
uncorrelated for all $\tau$.  The result of the fits to seven SFs are
shown on Fig.~\ref{fig:sf7}.  The values of the 5 parameters as a
function of the wavelengths are displayed in Fig.~\ref{fig:par}.

$p_0$ (Fig.~\ref{fig:par}a) is an experimental determination of the
uncertainty.  Comparing this parameter to the mean spectrum, one sees
that the relative uncertainty reached is between 2.7 and 4.6 \% with
IUE observations, and between 1.2 and 2.6 \% with the optical
observations.

$p_1$ (Fig.~\ref{fig:par}b), the end of the short-term variation, does
not show evidence of variation throughout the 29 light curves, with a
mean of 0.46 yr, and a very small dispersion of 0.054 yr. Note however
that, to achieve this result, the value of $p_1$ has been constrained
to be smaller than 1 yr in the 2 ``reddest'' IUE light curves
(otherwise, $p_1$ becomes larger than 10 yr). We adopted this
procedure, only because unconstrained fits to the optical SFs give
again values very close to 0.46 yr. It therefore appears to us that
the same plateau is identified in all the ultraviolet and optical
lightcurves, the more so as the samplings in the optical and
ultraviolet domains are completely different.

$p_2$ (Fig.~\ref{fig:par}c) is the amplitude of the \AAA\ term of the
SFs. It decreases slowly in the ultraviolet domain, but much more
rapidly in the visible domain; this is comparable to the variability
spectrum (Fig.~\ref{fig:sigma}b).

$p_3$ (Fig.~\ref{fig:par}d) is the slope of the \AAA\ term of the SFs.
The slope of the SF is related to that of the Fourier power spectrum:
The larger the slope of the SF, the smaller that of the Fourier power
spectrum.  $p_3$ decreases with wavelength, mostly in the optical
domain. This indicates that variability on short time scale is
comparatively more important in the optical domain than in the
ultraviolet domain. Caution must however be taken with this direct
interpretation. The shape of the \BBB\ term has been set arbitrarily,
and the value of $p_3$ may be affected by this choice.

$p_4$ (Fig.~\ref{fig:par}e) determines crudely the amplitude of a
longer-term variation (whose real SF is unknown). This \BBB\ term is
clearly needed in the optical domain, but is much less evident in the
ultraviolet domain, and its amplitude at these wavelengths is mostly
defined by a handful of values at the end of the ultraviolet SFs, and
thus not reliable. One sees however that its amplitude increases
dramatically above 2\,800 \AA, and joins with little discontinuity the
amplitude measured in the U band. This shows that the \BBB\ term
probably exists in the ultraviolet domain, although its intensity is
difficult to estimate.

\section{The two components of the \BB\ of \3c}
\label{sec:twocomp}
Eye inspection of the light curves, and the results from
Sect.~\ref{sec:tscal} give the impression that there are really two
differently varying components in \3c. We propose here a method to
decompose the light curves into these two tentative components.

\subsection{Decomposition method}
\label{sec:decomp}
PW96 have observed that the ultraviolet fluxes at two different
wavelengths are very well connected by an affine relationship (cf.\ 
Fig.~1 and the discussion in Sect.\ 2.2 \& 2.3 in PW96). They
therefore claimed that the \BB\ light curves could be the sum of two
components:
\begin{equation}
\label{eq-flin}
f(\nu,t)=\A\cdot \varphi(t)+\B,
\end{equation}
where $\A\cdot \varphi(t)$ is a variable component and $\B$ is a
stable component. $\varphi(t)$ is a function that determines the
complete temporal behaviour of the ultraviolet emission of the source.
It is unknown, but it can be approximated by a ``reference flux'', the
flux in a window located at the ``blue'' end of the IUE spectra
(around 1\,250 \AA). Using this reference flux, one can derive the
exact spectral shape of $\A$ (but not its normalization), and a set of
possible $\B[x]$ distribution, $0\!\leq\!x\!\leq\!1$, $x\!=\!0$
maximizing the contribution of $\A\cdot \varphi(t)$ at the frequency
of the reference flux, and $x\!=\!1$ minimizing it.

\begin{figure*}[p]
  \hspace*{-10mm}\mbox{\psfig{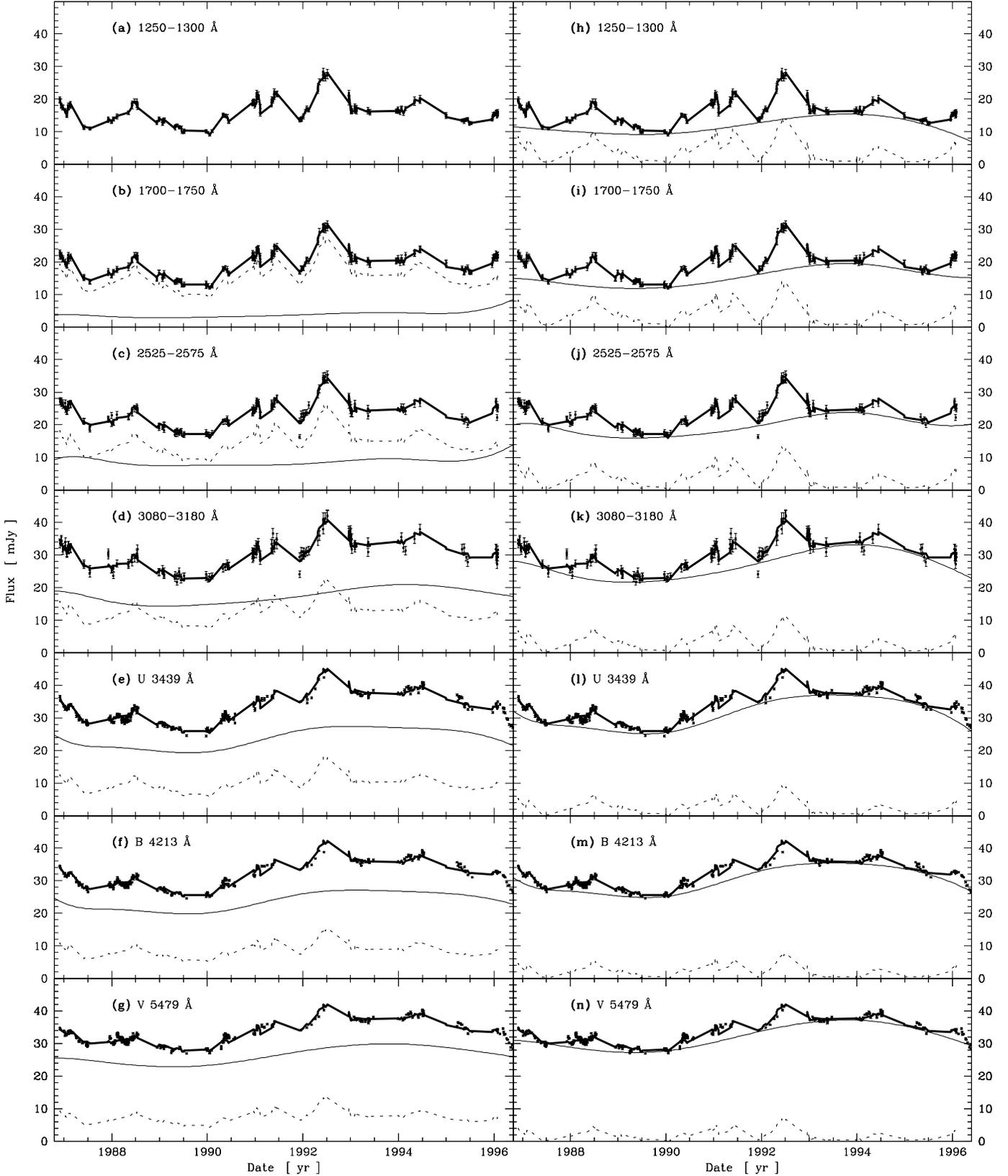}}
  \caption{{\bf a--g} Two-component fits to 7 light curves with the
    $\Bmin$ assumption. {\bf h--n} idem with the $\Bmax$ assumption.
    The thick solid line is the sum of the two components, the \AAA\ 
    component (dashed line), and the \BBB\ component (thin solid
    line).}
  \label{fig:comp7}
\end{figure*}

Sect.~\ref{sec:tscal} shows that Eq.~(\ref{eq-flin}) cannot explain
all the light curves. We test the possibility that what has been
called ``stable component'' in PW96 is actually variable.
Eq.~(\ref{eq-flin}) could be modified in the following way:
\begin{equation}
\label{eq-f2var}
f(\nu,t)=\A\cdot \varphi(t)+\B(t),
\end{equation}
Provided that $|\B(t_1)-\B(t_2)|\ll\A\cdot
|\varphi(t_1)-\varphi(t_2)|, \forall t_1,t_2$, Eq.~(\ref{eq-flin}) is
a sufficient approximation of Eq.~(\ref{eq-f2var}). This is why it
could be applied in PW96. With the increase of the time span of the
observations, and the addition of optical measurements, this condition
is not satisfied anymore. Hereafter we shall write ``\AAA'', and
``\BBB''\footnote{We call these components ``\AAA'' and ``\BBB'',
  because we shall see that the former is stronger at short, ``blue''
  wavelengths, and the latter at long, ``red'' wavelengths}
respectively the components $\A\cdot \varphi(t)$, and $\B(t)$ in
Eq.~(\ref{eq-f2var}). Their contributions to the SFs will be the \AAA\ 
and \BBB\ terms respectively.  We try to solve Eq.~(\ref{eq-f2var}) by
fitting the light curves with the sum of the \AAA\ and the \BBB\ 
components.  The \AAA\ component is given by the 1\,250--1\,300 \AA\ 
light curve and normalized by an unknown parameter. We have \emph{a
  priori} no information on the \BBB-component light curve; we
therefore define its light curve with 8 parameters, which are the
values of a cubic-spline function at abcissae 1984, 1986,\ldots, 1998
yr. The result of these fits are shown for seven light curves in
Fig.~\ref{fig:comp7}.  Note that the fits cannot be perfect in the
optical domain, because the value of $\varphi(t)$ must be interpolated
from the IUE observations.  The general agreement is however very
good.

As in PW96, the decomposition is not unique. Indeed the \BBB\ 
component may contribute at 1\,250 \AA\ also. The problem is actually
worse than in PW96, as it is not only a constant that is unknown, but
a whole light curve. We fix the maximum \BBB\ component by assuming
that it goes smoothly through the successive minima of the
1\,250-1\,300 \AA\ light curve (those around $t=1987.5,~ 1990,~ 1992,~
1993,~ 1993.5, 1994,~\mathrm{and}~ 1995$).  We shall call ``the
$\Bmin$ (resp.\ $\Bmax$) assumption'' the result of the decompositions
with the assumption of a negligible (resp.\ maximal) \BBB\ component
at 1\,250-1\,300 \AA.

\subsection{The \AAA\ component}
\label{sec:bba}
\begin{figure}[tb]
  \mbox{\psfig{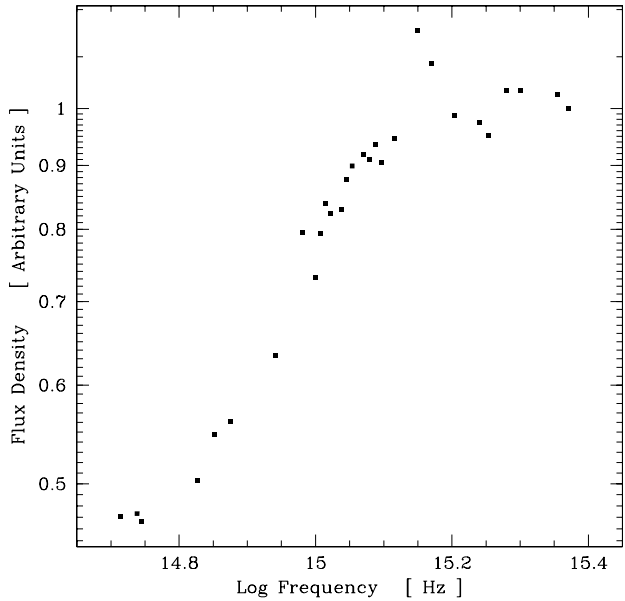}}
  \caption{Spectral energy distribution of the \AAA\ component
    normalized to its amplitude at 1\,250 \AA.}
  \label{fig:aa3c}
\end{figure}

The shape of the \AAA\ component is shown on Fig.~\ref{fig:aa3c}. As
in PW96, its normalization, but not its spectral shape, depends on the
assumption made on the \BBB\ component.  There is a clear curvature in
the ultraviolet-to-optical spectrum, and it is clearly not possible to
fit the entire spectrum with a single power-law. The absence of
discontinuity shows that this curvature cannot be due to an
instrumental problem.  A fit with two power-laws gives an index of
$-0.16$ at wavelengths shorter than 2\,600 \AA, and about $-1$ at
wavelengths longer than 2\,600 \AA\ ($f_\nu\sim\nu^{-\alpha}$).  Note
that the curvature in the \AAA\ spectral energy distribution has
already been observed for a subset of the data covering the years
before 1991, and where the \BBB\ component was considered constant
\cite[]{PW@96@bluevar}, and it is therefore not an artifact produced
by the introduction of a varying \BBB\ component.

\subsection{The \BBB\ component}
\label{sec:bbb}
We can see in Fig.~\ref{fig:comp7} that the smoothed light curves of
both the $\Bmin$ and $\Bmax$ components are very similar at all
wavelengths: they show a minimum around 1989--1990 and a maximum
around 1993-1994. It is important to remark that no \emph{a priori}
assumption on the light curve has been made in the $\Bmin$ case. The
similarity of the \BBB\ component at all the different frequencies is
an argument in favour of the reality of the decomposition.
\begin{figure}[tb]
  \mbox{\psfig{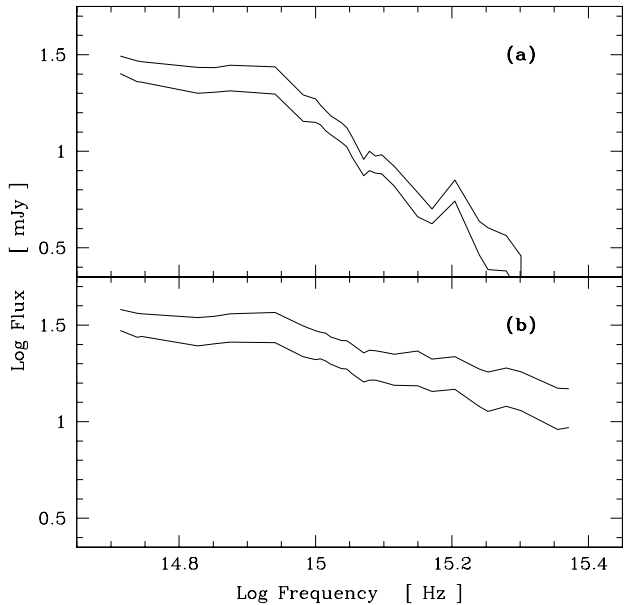}}
  \caption{Spectral energy distribution of the \BBB\ component.
    {\bf (a)} $\Bmin$.  {\bf (b)} $\Bmax$.  In each panel, upper
    curve: $t=1993.0$ (maximum flux); lower curve: $t=1989.0$ (minimum
    flux)}
  \label{fig:bb3c}
\end{figure}

The $\Bmin$ and $\Bmax$ spectral energy distributions at $t=1993.0$
and $t=1989.0$ (the extrema) are plotted on Fig.~\ref{fig:bb3c}.
There is still a strong flux below (in frequency) the Balmer edge,
which can be explained neither by emission lines, nor Balmer, and
Paschen continua, although they can modify slightly the shape of the
\BBB\ component (in particular, the small bump around $10^{14.9}$ Hz
can be explained by Balmer continuum and line emissions).  The $\Bmin$
component deviates strongly from a unique power-law, while the $\Bmax$
component is compatible with a single power-law of index about $0.7$.

\section{Delay between the light curves}
\label{sec:lag}
\begin{figure}[tb]
  \mbox{\psfig{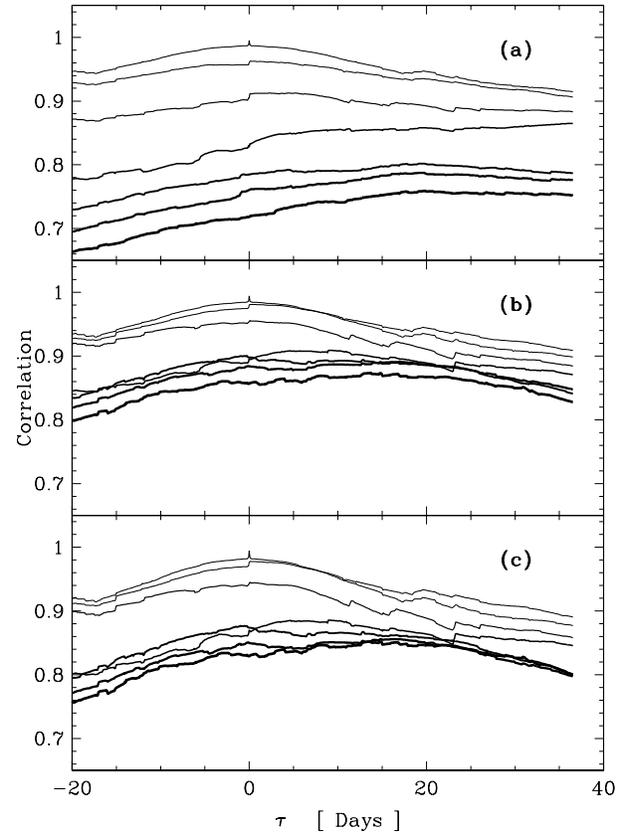}}
  \caption{Correlations between the IUE 1\,250-1\,300 \AA\ light curve
    and 7 IUE and optical light curves (the thicker the line, the
    larger the wavelength of the second light curve).  {\bf (a)} Total
    light curves. {\bf (b)} \AAA\ with the $\Bmin$ assumption.  {\bf
      (c)} \AAA\ with the $\Bmax$ assumption}
  \label{fig:smallc}
\end{figure}

Fig.~\ref{fig:correl} already indicates that the correlation peak
close to $\tau=0$ has an amplitude that decreases and a centroid that
moves slightly towards larger values of $\tau$ as the wavelength of
the second light curve increases. Fig.~\ref{fig:smallc} emphasizes
this effect with a closer look at the central peak of the same
correlations for the complete light curves, and for the \AAA\ 
components obtained with the $\Bmin$ and $\Bmax$ assumptions
respectively. This is an evidence for a delay between the different
light curves.
\begin{table*}[htb]
  \caption{Results of the ``$\tau$'' test. ``Probability'' is the
    probability that the test performed on uncorrelated parent
    populations gives a value of $\tau$ smaller or equal
    to the measured one}
  \label{tab:lag}
  \begin{tabular}{cccccccccc}
    \hline
    \rule{0pt}{1.2em}Light curves &\multicolumn{3}{c}{Total}&\multicolumn{3}{c}{\AAA\ with $\Bmin$}&
    \multicolumn{3}{c}{\AAA\ with $\Bmax$}\\
    &$t_2$ (days)&$\tau$&Probability&$t_2$
    (days)&$\tau$&Probability&$t_2$ (days)&
    $\tau$&Probability\\
    \hline
    \rule{0pt}{1.2em}All&$13.8$&$-0.79$&$0^*$&$\z4.2$&
    $-0.66$&$0^*$
    &$\z4.4$&$-0.66$&$0^*$\\
    SWP &&$-0.90$&$1.0\,10^{-5}$&&$-0.33$&$9.4\,10^{-2}$&&$-0.33$&$9.4\,10^{-2}$\\
    LWP/LWR &&$-0.44$&$2.5\,10^{-6}$&&$-0.56$&$3.0\,10^{-9}$&&$-0.56$&$3.0\,10^{-9}$\\
    Optical &&$-0.71$&$7.4\,10^{-4}$&&$-1.00$&$4.8\,10^{-7}$&&$-0.90$&$1.0\,10^{-5}$\\
    \hline
  \end{tabular}\\
\noindent$^*$These probabilities are formally smaller than $10^{-40}$
\end{table*}

In view of Fig.~\ref{fig:smallc}, we would like to establish the
statistical significance of the delay, and, assuming that it is real,
estimate its amplitude. Note that the decomposition is still
valid in spite of the delay, because the value of the correlation at
$\tau=0$ is very close to its maximum.  The lags have been estimated
using cross-correlation analysis (as in Sect.~\ref{sec:corr}).  More
sophisticated methods have been tried, but they gave less reliable
results. The $\chi^2$ method of \cite{PressAl@92@optlag} is not
appropriate, because the covariance matrix is dominated by measurement
uncertainties, and thus badly known, at very short lags, where we
expect to find the delay. The ``dispersion spectrum'' method of
\cite{PeltAl@94@delay1} suffers from a similar problem, namely that,
because of the short lags investigated, most of the pairs of
observations used in the calculation of the ``dispersion spectrum''
are separated by too long a delay. We note however that the
qualitative results obtained with these two methods are compatible
with those obtained with the ICCF.

\begin{figure}[tb]
  \mbox{\psfig{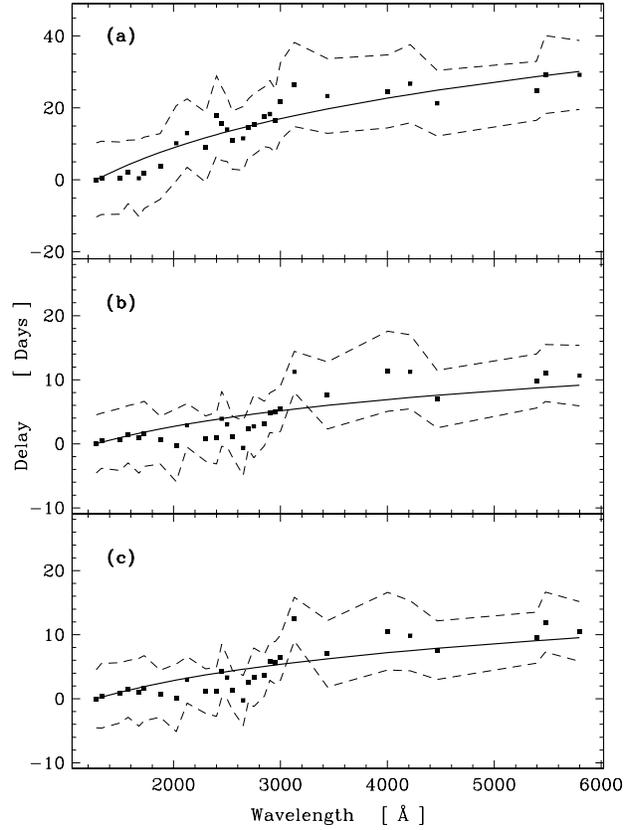}}
  \caption{Delay with respect to the first light curve. The points are
    the delays averaged on the 29 measurements, and the dashed lines
    show their 3$\sigma$ dispersion. The thick line is the fitted
    delay.  {\bf (a)} Total light curves.  {\bf (b)} \AAA\ with the
    $\Bmin$ assumption.  {\bf (c)} \AAA\ with the $\Bmax$ assumption}
  \label{fig:delay}
\end{figure}

We define a procedure to estimate the delay between two light curves
using the ICCF method: First we calculate the $29\times 29$ ICCFs
between all the pairs of light curves for $-0.1\!\le\! \tau\!\le\!
+0.1$ yr; then we fitted a 2nd-degree polynomial to the ICCF, whose
abscissa of the maximum defines the delay. We did this for the
observed light curves, the \AAA\ component with the $\Bmin$
assumption, and the \AAA\ component with the $\Bmax$ assumption. We
then calculate the average delay $\tilde{\Delta}_i$ between the
first and the $i^{\mathrm{th}}$ light curve (1\,250--1\,300 \AA) with
the relationship:
\begin{equation}
  \label{eq-delay}
  \tilde{\Delta}_i=\frac{1}{n}~ \sum_{j=1}^n~ \Delta_{1,j}+\Delta_{j,i},
\end{equation}
where $n=29$ is the number of light curves, and $\Delta_{k,\ell}$ is
the delay between the $k^{\mathrm{th}}$ and the $\ell^{\mathrm{th}}$
light curves. A positive delay means that the first light curve
precedes the seconds one. This average has the purpose to decrease the
effect of the uncertainties on the lags.  The $\tilde{\Delta}_i$ delay
curves are shown in Fig.~\ref{fig:delay}, together with the dispersion
of the individual delays used in the calculation of
$\tilde{\Delta}_i$.
The delays can be approximated by the function: $\tilde{\Delta}_i
\simeq t_2\cdot \log_2( \lambda_i/1\,275 \mbox{~\AA})$. The parameter
$t_2$ gives the delay observed between two light curves at wavelengths
$\lambda_0$ and $2\cdot \lambda_0$ respectively. Their values in the
three cases are reported in Table~\ref{tab:lag}. No difference can be
observed between the $\Bmin$ and $\Bmax$ decomposition, with a $t_2$
of about 4 days, but the $t_2$ obtained without decomposition is 3
times larger. This could be due to a contamination by the \BBB\ 
component (see Sect.~\ref{sec:aabb}).

The global significance of the existence of a delay can be tested by a
modified Kendall's $\tau$ test \cite[e.g.][]{PressAl@92@NUMREC}. The
original test (and the version used below) has the very interesting
property to be completely non-parametric. Assuming a set of data
$(x_i;y_i),~ i=1,\ldots,n$, the $\tau$ test actually measures the
excess of \emph{``concordant''} (or alternatively
\emph{``discordant''}) pairs of data taking into account all the
possible pairs, a pair of data $[(x_i;y_i);(x_j;y_j)]$ being called
\emph{``concordant''} (respectively \emph{``discordant''}) if either
($x_i\!>\!x_j$ and $y_i\!>\!y_j$), or ($x_i\!<\!x_j$ and
$y_i\!<\!y_j$) (respectively, if either ($x_i\!<\!x_j$ and
$y_i\!>\!y_j$), or ($x_i\!>\!x_j$ and $y_i\!<\!y_j$)). We cannot use
directly this scheme, as a delay is not defined for one wavelength,
but only for a couple of wavelengths.  We therefore replace the
condition $y_i\!>\!y_j$ (for instance) by the condition
$\Delta_{i,j}\!>\!0$.  Kendall's $\tau$ can then be estimated using
Eq.~(14.6.8) of \cite{PressAl@92@NUMREC}. The probability to obtain a
given value for $\tau$ from uncorrelated parent populations (which
means in this case that their lags are random) can be calculated from
simple combinatorics arguments.

Table~\ref{tab:lag} gives the correlation coefficients and the
probability that the test performed on two uncorrelated populations
gives a value of $\tau$ smaller or equal to the measured one.  The
minus sign indicates that the long-wavelength light curves follow the
short-wavelength ones, compatible with Fig.~\ref{fig:delay}. If one
uses the 29 light curves, it is absolutely impossible that the
correlation coefficients are obtained by chance. However it is
possible that the correlation is due to the sampling. To exclude this
possibility, we have applied the $\tau$ test to light curves that have
exactly the same sampling. Here also, the probabilities that the
correlation coefficients are obtained by chance are completely
negligible. As the tests are independent, the probabilities to obtain
these correlation coefficients are smaller than $10^{-15}$ both for
the complete light curves and those of the \AAA\ components. Thus we
conclude that the existence of a lag is firmly established.

\section{Discussion}
\label{sec:disc}

\subsection{Is the decomposition physical?}
The similarity of the central peaks in all the correlations and the
important differences in the structure functions offer a very
compelling argument in favour of the physical existence of two
separate \AAA\ and \BBB\ components.  They imply that the long term
variability is not the result of ``dilution'' of the light curve seen
at short wavelength by a larger emission size or a longer cooling time
scale, as the central peak would have appeared more and more smeared
when the wavelength increases.  The absence of smearing is confirmed
by the structure function analysis, which shows that the $p_1$
parameter (see Eq.~(\ref{equ:funcsf})) is constant.  The fact that the
``birth'' of the \BBB\ component can be seen in the rapid increase of
$p_4$ around 2\,800 \AA\ is a strong argument against an instrumental
origin of this component.

The light curves of the \BBB\ components were completely free
parameters in the fits of Sect.~\ref{sec:decomp}. In spite of this,
the light curves of the $\Bmin$ and $\Bmax$ components are all very
similar, and the resulting spectrum (Fig.~\ref{fig:bb3c}) is smooth
and quasi-monotonical.  The near-infrared light curves
\cite[]{TPCAl@98@3cdata} are similar to the one of the \BBB\ 
component, while no variation in the form of the \AAA\ component is
detectable at these wavelengths.  Because of its spectral shape, the
near-infrared emission of \3c\ is usually attributed to thermal
emission from hot dust \cite[]{Barvainis@87@pouss}.  This component
has a negligible contribution in the optical domain, because dust is
destroyed at temperature above $\sim$1\,500 K, and should also be
constant on time scales of the order of the year, as the size of the
emitting region is about 10 pc \cite[]{Barvainis@87@pouss}. Thus we
conclude that another variable component contributes to the
near-infrared emission of \3c, and that it is very probably the \BBB\ 
component, owing to the similarity of the light curves.

Taking these points into account, we consider that the mathematical
decomposition of Eq.~(\ref{eq-f2var}) succeeds in isolating two
independent physical components present in both the ultraviolet and
optical light curves of \3c.

\subsection{The lag and the hump in the correlations}
\label{sec:aabb}

\begin{figure}[tb]
  \mbox{\psfig{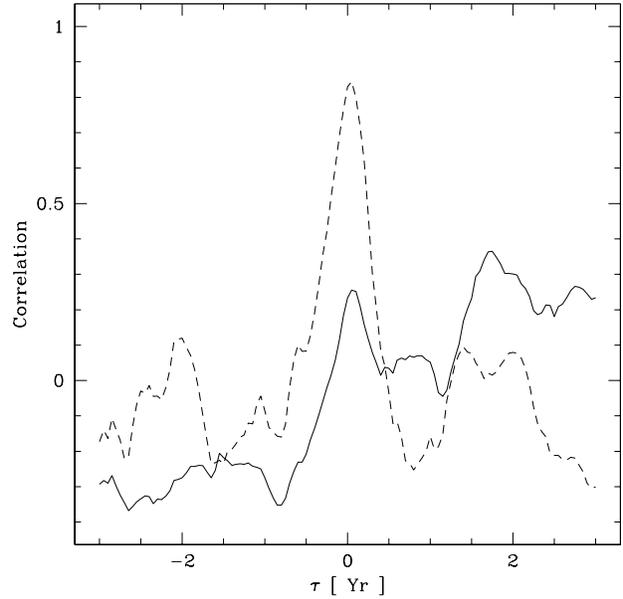}}
  \caption{ Correlation between the \AAA\ component at 1\,250-1\,300
    \AA\ and respectively the optical V light curve (solid line) and
    the \AAA\ component in the optical V band (dashed line).}
  \label{fig:correl2}
\end{figure}

We observed the presence of a hump between 0 and 2 yr in the
correlations shown in Fig.~\ref{fig:correl}.  On
Fig.~\ref{fig:correl2}, we have plotted the correlation between the
\AAA\ component with the $\Bmax$ assumption at 1\,250-1\,300 \AA\ and
the optical V light curve (solid line). Again, a similar structure is
found. If one calculates the correlation between the \AAA\ component
at 1\,250-1\,300 \AA\ and the \AAA\ component in the optical V light
curve (dashed line), the hump is strongly suppressed.  We conclude
that the hump appears whenever a light curve containing the \AAA\ 
component is correlated to another one containing the \BBB\ component.
A much longer monitoring would be necessary to determine whether this
feature is real, or due to some soincidence in the light curves. If
its reality is confirmed, it would have important consequences on the
interpretations of the \AAA\ and \BBB\ components, as it would suggest
a physical connection between the two components.

The hump may affect the shape the central peak, and hence its
centroid. This cannot fully explain the lags that we have found, as we
also observe a lag when the \BBB\ component is removed. It can however
explain why a longer lag is observed when the \BBB\ component is not
removed. The value of the lag is unchanged, whether we choose the
$\Bmin$ or the $\Bmax$ assumption. It means that the influence of the
\BBB\ component is negligible even with the $\Bmin$ assumption.  We
therefore conclude that the lag is a property of the \AAA\ component,
and that the correct amplitude of the lag is $t_2\simeq 4$ days.

It is the second time that a lag between the different light curves of
the \BB\ of an AGN is found. The first object is NGC~7469, where a lag
has been detected both within the IUE SWP wavelength range
\cite[]{WandersAl@97@7469laguv}, and between IUE and optical light
curves \cite[]{CollierAl@98@7469lagopt}. The value of $t_2$ in
NGC~7469 is 0.6 day. The average ultraviolet luminosity of NGC~7469 at
1\,300\AA\ is about $1.5\,10^{27}$ erg s$^{-1}$ Hz$^{-1}$, 1000 times
less luminous than \3c\ ($1.25\,10^{30}$ erg s$^{-1}$ Hz$^{-1}$).  It
suggests that the lag may increase with $L_{1250~
  \mbox{\scriptsize\AA}}^{1/3}$.

\subsection{The nature of the \AAA\ component}
\label{sec:natbba}

\subsubsection{Time-series properties}
\label{sec:varbba}
Since the broadness of the Fourier transform of AGN light curves has
been demonstrated (to our knowledge by \cite{Kunkel@67@broadfour}),
many works have stated that the Fourier power-spectra of AGN light
curves follow a power-law.  From the $p_3$ parameter we can estimate
that the index of the power-law would be about $-2.5$, which is, for
instance, too steep to be due to a random-walk process. It must be
realized however that almost any poorly-known broad Fourier
power-spectra will appear power-law-shaped. Thus we do not want to
discuss more the power-law index deduced from the $p_3$ parameter.

The $p_1$ parameter is the signature that there exists a frequency
below which the Fourier power spectrum becomes negligible. Comparison
with SFs of simulated light curves shows that this frequency is about
0.65 yr$^{-1}$. This means that the system that emits the \AAA\ 
component has no ``memory'' of the state that it had 2 yr before. It
is not obvious to find a physical continuous system that explains this
low-frequency cut-off.  However, the discrete-event models
\cite[]{PC@97@event} naturally incorporate such a feature. This is
discussed in Sect.~\ref{sec:event}.

The Fourier power-spectrum of a light curve has a high-frequency limit
due to the size of the source (in the direction perpendicular to the
observer).  From the V light curve (which has the smallest
uncertainties), a lower limit on the high-frequency limit can be put
to $\sim$(10 day)$^{-1}$. Crossed at the speed of light, it
corresponds to about 100 Schwarzschild radii (R$_\mathrm{S}$) of a
black hole with a mass $10^9$ M$_\odot$. This is only the limit size
parallel to the observer's direction. In particular, as \3c's jet is
seen with an angle of about $10^\circ$ (as deduced from the presence
of superluminal motions), an accretion disk perpendicular to the jet
could be much larger.

\subsubsection{A discrete-event model interpretation of the variability}
\label{sec:event}
We explore the possibility that the variability of the \AAA\ component
is due to events occurring at random epochs. The main reason to make
this assumption is that the \AAA\ component would naturally exhibit a
maximum variability time scale, as soon as the events have a finite
life time. It can indeed be shown that the SF of a superimposition of
an arbitrary number of similar events is proportional to the SF
restricted to one single event \cite[]{P@96@these}. Thus, the SF of
the total light curve will be constant for all $\tau$ larger than the
event life time.

\begin{figure}[tb]
  \centerline{\mbox{\psfig{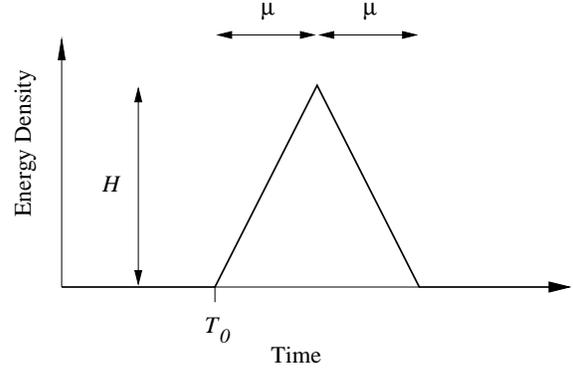}}}
  \caption{Light curve and parameters of the events used in
    Sect.~\protect\ref{sec:event}}
  \label{fig:triangle}
\end{figure}
We want to check whether all the observed variability properties of
the \AAA\ component can be reproduced. Let us use a very simple event
shape, namely a triangular one, shown in Fig.~\ref{fig:triangle}. Its
parameters are the time of birth of the event, $T_0$ (expressed in yr;
this is the only parameter allowed to be different for each event),
the event amplitude, $H$ (expressed in erg s$^{-1}$ Hz$^{-1}$), and
the time needed to reach the maximum flux, $\mu$ (expressed in yr). A
further parameter is the average event rate $\overline{N}$ (expressed
in number of events per yr). The light curve properties that have to
be matched are: the mean flux, the observed variance, and the location
of the plateau in the SF. In principle the shape of the increasing
part of the SF must also be matched, but the SF is not very sensitive
to changes in the event shape, and practically any reasonable event
shape fits the SF. The location of the plateau is only a function of
$\mu$, and it corresponds to the one found in the SF of the \AAA\ 
component if $\mu\simeq 0.4$ yr (this means that the total duration of
an event is $0.8$ yr).  From Eqs~(11) and (B2) of \cite{PC@97@event},
we find the following relationships for the mean luminosity and its
variance (assuming that the $\Bmax$ assumption is correct):
\begin{equation}
  \label{eq-parevent}
  \left\{ ~
    \begin{array}{llll}
      \overline{L_{1250/(1+z)~\mbox{\scriptsize\AA}}}&=&\overline{N}\cdot H\cdot
      \mu&=\\
      &=&3.32\,10^{29}\mathrm{~erg~s}^{-1}\mathrm{~Hz}^{-1}\\[2mm]
      {\mathrm{Var}}(L_{1250/(1+z)~\mbox{\scriptsize\AA}})&=&\overline{N}\cdot
      H^2\cdot \frac{2}{3}\mu&=\\
      &=&5.99\,10^{58}\mathrm{~(erg~s}^{-1}\mathrm{~Hz}^{-1})^2\\
    \end{array}
  \right.
\end{equation}
where $z=0.158$ is the redshift. The mean luminosity and the variance
of the \AAA\ component are calculated using a Hubble constant $H_0=60$
km s$^{-1}$ Mpc$^{-1}$, and a deceleration parameter $q_0=0.5$. We can
solve Eq.~(\ref{eq-parevent}), which gives:
\begin{equation}
  \label{eq-solvpar2}
  \left\{ ~
    \begin{array}{llrl}
      \overline{N}&=&3.07&\mathrm{~events/yr}\\
      H&=&2.7\,10^{29}&\mathrm{~erg~s}^{-1}\mathrm{~Hz}^{-1}\\
      \mu&=&0.4&\mathrm{~yr}\\
    \end{array}
  \right.
\end{equation}
Using $\Bmin$, the event luminosity is slightly smaller, and 30 events
per year are required. Assuming that the event shape is the same at
all wavelengths, Fig.~\ref{fig:aa3c} allows us to calculate $H$ as a
function of the frequency.  We deduce that the total event energy
between 1\,200 and 6\,000 \AA\ (observed wavelengths) is $3\,
10^{51}$ erg, if the emission is isotropic. We can therefore rule out
models where the events are supernova explosions
\cite[]{Terlevich@92@agnsn} for energetic reason, as it is 100 times
larger than the luminous energy of a supernova.  If the \AAA\ 
component is responsible for the soft-excess component, as suggested
by \cite{WFink@93@bbsx}, and extends up to 100 eV, the total event
energy becomes of the order of $10^{52}$ erg.  We note that using
different event shapes would change the details of the figures, but
not their orders of magnitude.

\cite{CPW@96@coll} have proposed that the luminosity originates from
the kinetic energy of colliding stars. Fixing the black hole mass to
$10^9$ M$_\odot$, it is possible to reproduce satisfactorily the
variability properties of the \AAA\ component, if the central stellar
density is $10^{12}$ stars pc$^{-3}$, and if the core radius is 100
\RS\ ($\equiv 10^{-2}$ pc, or 10 light-days) in the $\Bmin$ case, and
40 \RS\ in the $\Bmax$ case.  This cluster would contain about $10^6$
stars inside its core radius.

The magnetic blobs introduced by \cite{GaleevAl@79@blob} and used by
\cite{HaardtAl@94@blob} are another kind of event candidates.  To
store $10^{52}$ erg in the form of magnetic energy with a field of
$10^5$ G requires a volume of the order of $10^{40}$ cm$^3$, i.e.\ a
cube with a side of $10^{13}$ cm (0.1 \RS). To check the plausibility
of these values would require a detailed physical treatment of the
phenomenon, which is out of the scope of this work.

\subsubsection{Synchrotron radiation model of the \AAA\ component}
Synchrotron emission has been considered as a candidate for the \BB\ 
emission. To obtain a ultraviolet spectral energy distribution
comparable to the one of the \AAA\ component, a synchrotron emitting
medium requires a very flat electron energy distribution, with an
index not smaller than $\sim -1.3$. The optical part of the \AAA\ 
component requires then that the medium becomes self-absorbed around
$\nu=10^{15}$ Hz. Using the formulae of
\cite{RybickiLightman@79@RadProc}, we find that, if the electron
energy distribution is given by $N(\gamma)\propto \gamma^{-1.2}$, and
if $B=1$ G, it requires the size of the emitting medium to be of the
order of $10^8$ cm, and an electron density of the order of $10^{23}$
cm$^{-3}$. The size depends little on the magnetic field, the
electron energy distribution index, and even Doppler boosting. The
required electron density is still $\sim 10^{21}$ cm$^{-3}$ if the
magnetic field is fixed to 1 T. The Thompson opacity of this medium is
extremely high ($>10^6$), which means that inverse Compton cooling
must strongly dominate synchrotron radiation, and thus make the
ultraviolet emission vanish, which rules out synchrotron radiation as
the explanation of the \BB\ of \3c.

\subsubsection{Bremsstrahlung radiation model of the \AAA\ component}
Optically-thin thermal bremsstrahlung emission already had some
difficulty in fitting the $\A$ component in \3c\ (PW96), as the
emission from such a mechanism has a spectral index of $\sim$0.2 when
$\mathrm{h}\nu\!\ll\!  \mathrm{k_B}T$ \cite[e.g.,
][]{GronenschildMewe@78@thin,Barvainis@93@free}, where h is Planck's
constant, $\mathrm{k_B}$ Boltzmann's constant and $T$ the temperature
of the optically thin medium. It follows from this property that a
flux distribution entirely due to optically thin emission cannot have
parts of its spectrum with negative spectral index, whatever the
temperature distribution of the emitting medium.  The spectral index
of the \AAA\ component in the optical domain is clearly negative,
which rules out the possibility of this emission mechanism in this
object.
  
On the other hand, thermal bremsstrahlung emission can be partially,
or even fully, optically thick. In that case, the thicker the emitting
medium, the closer the emitted spectrum to the one of a blackbody. We
calculated the spectrum of thermal bremsstrahlung emission for a
plane-parallel geometry using the formulae of
\cite{RybickiLightman@79@RadProc}. The \AAA\ distribution can be very
well fitted with such a model. If one imposes a temperature of
$2.73\, 10^6$ K (which is the temperature obtained by
\cite{WOCAl@94@simultxs1} with a fit on ultraviolet and soft X-ray
simultaneous observations) in the whole emitting medium, the electron
density is $9\, 10^{13}$ cm$^{-3}$, if the geometrical depth is
$10^{13}$ cm. The emitting medium becomes optically thick
($\tau_\nu=1$) below $9.5\, 10^{14}$ Hz ($\simeq$ 3150 \AA).

\cite{NayakshinMelia@97@bbbflares} have developed an interesting model
where bremsstrahlung emission is expected: An X-ray flare produced by
the discharge of a magnetic blob (as described by
\cite{GaleevAl@79@blob}) above the corona of an accretion disk heats
and compresses the corona, which cools down by emitting thermal
bremsstrahlung radiation.  With the above parameters, the spectrum of
the \AAA\ component is very well explained.  A possible difficulty is
that the condition $\tau_\nu\simeq 1$ close to $10^{15}$ Hz may
require a fine tuning, which must be realized for all events.  It must
be noted that, with these parameters, the corona is optically thick to
Thompson scattering ($\tau_{\mathrm T}\sim 600$). The system is
actually very similar to the one studied in detail by
\cite{CollinAl@96@spher}, which has been found to be the only viable
explanation of the emission of Seyfert galaxies using optically thin
clouds. A first consequence is that the resulting spectrum is modified
in a complicated way. We cannot however exclude that another set of
parameters produces the correct spectral energy distribution. Another
consequence is that the emitting medium cannot be instantaneously
heated by the X-ray radiation.  This can however be the cause of the
lag, and only a complicated time-dependent simulation of the physical
event can give credit to or exclude this model. This model also
predicts that soft X-ray emission must be dominated by bound-free and
bound-bound transitions. However we feel that the shape of the
soft-excess is not yet sufficiently constrained to reject this
possibility. The predicted relationship between the ultraviolet and
hard X-ray emissions matches well the observations
\cite[]{PCTW@98@UVXcorRome}.  The variability properties of the \AAA\ 
component finds here an interesting explanation in the context of
discrete-event models, if it is possible to store some $10^{52}$ erg
in magnetic blobs.

\subsubsection{Optically thick reprocessing model of the \AAA\ component}
\label{sec:repra}
Reprocessing on an accretion disk is the other alternative that has
been considered in PW96. We shall not discuss viscously-heated
accretion disk models, as they fail to explain the quasi-simultaneity
of the ultraviolet and optical variations \cite[]{CClavel@91@obscons}.
Even very simple models, where a disk is illuminated by an X-ray
source located above its centre and radiates locally as a black-body
whose temperature depends on the X-ray illumination, are capable of
explaining the range of power-law indices in the ultraviolet. The
presence of a spectral break can be easily accounted for, provided
that the disk has a finite outer radius.  More generally, a model that
uses a distribution of black bodies must not contain matter at
temperatures below $\sim$20\,000 K.\@ \cite{MattAl@93@reproc} have
calculated the X-ray and far-ultraviolet spectrum of a much more
realistic model of irradiated accretion disk including radiative
transfer and self-consistent vertical structure; unfortunately they
limit their calculations at wavelengths shorter than 1200 \AA.
However, it seems unplausible that the spectral break can be explained
without using a finite disk.  Indeed, in the model of
\cite{RossAl@92@viscdisk} (on which the model of
\cite{MattAl@93@reproc} is based) the spectral break appears because
the size of the accretion disk is limited to 50 \RS.\@ Using our
simplistic model, and assuming that the central black hole has a mass
of $10^9$ M$_{\odot}$, the correct distribution of the \AAA\ component
can be obtained if the disk has an outer radius of 0.03 pc$\equiv$ 300
\RS, and the X-ray source is located 1.5 mpc$\equiv$ 15 \RS\ above the
black hole.
  
A reprocessing geometry naturally accounts for the existence of a lag,
because colder parts of the reprocessing zone are located farther from
the hard X-ray source.  It means that the distance between the
ultraviolet emission zone and the optical emission zone is of the
order of 10 light-days ($\sim$ 100 \RS\ of a $10^9$ M$_\odot$ black
hole), if crossed at the speed of light.  The disk parameters
discussed here induce a delay of the order of 10 days between the
1\,250 \AA\ and the 5\,500 \AA, amazingly close to the observed value.

\begin{figure}[tb]
  \centerline{\mbox{\psfig{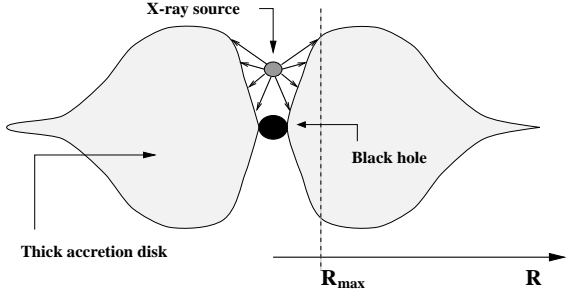}}}
  \caption{Sketch of an irradiated geometrically-thick accretion
    disk. Reprocessing does not occur at $R>R_{\mathrm{max}}$}
  \label{fig:thick}
\end{figure}

On the other hand, this model has major difficulties. The disk
parameters found here produce a disk much too cold to explain the
soft-excess emission, which is thought of to be the same component as
the \BB\ \cite[]{WFink@93@bbsx}.  Reprocessing models also predict an
iron line around 6.5 keV, which is not always present
\cite[]{YaqoobAl@94@asca3c}, and a reflection hump in the hard X-ray
spectrum, which has never been firmly detected
\cite[]{MaisackAl@92@hexeref}. These problems can be alleviated by
considering a geometrically-thick accretion disk
(Fig.~\ref{fig:thick}).  \cite{Madau@88@thick} has shown that the
X-ray spectrum of such a disk can be strongly modified by the effect
of re-absorption of photons emitted inside the funnel. It consideraby
increases the temperature of the accretion disk close to the black
hole, and the reflection features are attenuated. As a bonus, the disk
needs not to be truncated, as a maximum reprocessing size is naturally
given by the geometry. This size ($\sim$100 \RS) is moreover of the
order of magnitude predicted by the models \cite[]{Madau@88@thick}.
However, \cite{PCTW@98@UVXcorRome} did not find any correlation at
short lag between the ultraviolet and hard X-ray emission, although it
is predicted by reprocessing models.  Thus these models require that
the X-ray emission observed on Earth is different from the one seen by
the disk.

\subsection{The nature of the \BBB\ component}
\label{sec:bbblong}
Whether we take the $\Bmin$ or the $\Bmax$ assumption, the shape of
the \BBB\ component is very different. The spectral index of its
ultraviolet part increases from 0.8 to 2.7, the softest spectrum being
obtained with the $\Bmin$ assumption.  Whatever its normalization, the
light curve of the \BBB\ component is similar to the one shown in
Fig.~\ref{fig:comp7}.  The large range of possible spectral shapes
prevents us from obtaining strong evidence for any particular models
from the spectral shape alone. We therefore have to find additional
clues from the other properties of \3c.

The BATSE instrument aboard the Compton-GRO satellite monitors
quasi-continuously \3c\ in the 20-200 keV domain since 1991.  The
light curve can be seen in Fig.~7 of \cite{TPCAl@98@3cdata}; it shows
similarities with the one of the \BBB\ component: There is a maximum
in 1994, and variability on time scales of a few years (obviously
limited by the ``youth'' of BATSE) dominates short-time-scale
variability. We shall give two tentative interpretations on the basis
of this observation.

\subsubsection{Reprocessing of X-ray radiation on cold matter}
The similarity of the light curves of the \BBB\ component allows us to
exclude a viscously-heated accretion disk as the origin of the \BBB\ 
component.  However, reprocessing of hard X-ray radiation on an
accretion disk can explain both this similarity, and the correlation
between the \BBB\ component light curve and the BATSE light curve. As
an iron line has been detected around 6.5 keV
\cite[]{TurnerAl@90@xray3c}, we know that some reprocessing on cold
material must take place in \3c, and the \BBB\ component seems to be a
very good candidate. On the other hand, the absence of reflection hump
\cite[]{MaisackAl@92@hexeref} shows that the reprocessed radiation
cannot be very important.

The ``standard'' reprocessing picture already invoked for the \AAA\ 
component (see Sect.~\ref{sec:repra}) can account for the spectral
energy distribution of the $\Bmax$ component, if an X-ray source of
about $10^{47.5}$ ergs s$^{-1}$ is located at 60 \RS\ above the disk,
assuming that the mass of the central black hole is $10^9$ M$_\odot$.
The disk need not be truncated. This model has however an important
drawback. The variability of the \BBB\ component is larger (in term of
flux difference) in the infrared domain (J, H, and K bands) than in
the optical and ultraviolet domains, while the opposite is expected
from the above model; moreover, the disk becomes very large, and
variability should not be simultaneous anymore.

\subsubsection{The hidden blazar}
\3c\ is usually classified as a blazar, because of its radio and
gamma-ray properties.  Actually its spectral distribution below 100
$\mu$m (and, although to a somehow lesser extent, above 10 keV) is
very close (both in spectral shape and in intensity) to the one of
\3c's close friend, 3C~279 (cf.\ Fig.~8 from
\cite{VonMontignyAl@97@mw3cblazar} and Fig.~1 from
\cite{MaraschiAl@94@mw279}).  \3c\ however largely dominates 3C~279 in
the \BB\ domain, but their spectral shapes are so different that there
is little doubt that one observes different components in both
objects.  One can therefore easily admit that the component that emits
optical radiation in 3C~279 should also be present in \3c. Thus the
\BBB\ component is possibly the signature of the ``blazar \3c'' in the
optical-to-ultraviolet domain.

Models of blazar spectral energy distribution -- like synchrotron
self-Compton (SSC), external Compton, and proton-initiated cascade --
predict a synchrotron radiation component in the optical emission.
Therefore, the \BBB\ component -- at least in the SSC model -- should
be correlated to the hard X-ray emission. The synchrotron component in
\3c\ should be much more important than in 3C~279 to account for the
\BBB\ component, which appears quite possible considering the numerous
free parameters in these models.  There is actually a very strong
evidence that synchrotron radiation contributes to the optical
emission of \3c: the optical flux can have a polarization as high as
2.5 \% \cite[]{DeDiegoAl@92@3coptpol}. Moreover the polarized spectrum
is flat in the optical domain, which corresponds well to the \BBB\ 
spectral energy distribution.  If the \BBB\ component is emitted by
synchrotron radiation, the extension of the \BBB\ component down to
the infrared domain is expected, and the large amplitude of the \BBB\ 
component variability is not a problem anymore.

\section{Summary and possible scenario}
\label{sec:conc}
As a result of a time series analysis of 29 optical and ultraviolet
light curves of \3c, we arrive at the conclusion that two components,
written here \AAA\ (for ``Blue'') and \BBB\ (for ``Red''), contribute
to the optical and ultraviolet emission. While the relative strength
of these components cannot be accurately determined, the \BBB\ 
component dominates at optical wavelengths, and the relative
contribution of the \AAA\ component increases quickly towards short
wavelengths.

The \AAA\ component has the interesting property of varying only on
time scales shorter than 2 years, independently of the wavelength. If
this feature is interpreted in terms of discrete-event models, which
would naturally explain the presence of a low-frequency cut-off, the
event rate is of the order of 10 per years, and each of them liberates
some $10^{52}$ erg, if this component is also responsible for the
soft-excess emission. The spectrum of the \AAA\ component including
the soft-excess emission is best explained by thermal bremsstrahlung
emission, provided that the emitting medium becomes optically thick
around $10^{15}$ Hz, and by reprocessing on a (probably
geometrically-thick) accretion disk. The fine tuning of the free-free
opacity may however be an argument against thermal bremsstrahlung
emission, as is the difficulty of instantaneously radiatively heating
a zone with a high Thompson scattering opacity. The presence or
absence of bound-free and bound-bound transitions in the soft X-ray
emission of \3c\ is a fundamental test within the capabilities of the
XMM and AXAF telescopes. We find also that the longer wavelength light
curves follow those at shorter wavelengths with a delay. Its amplitude
is of the order of 4 days for two light curves whose wavelength ratio
is 2.  This is very well accounted for by reprocessing on an accretion
disk. The explanation is less obvious in the case of thermal
bremsstrahlung, but it might be due to the heating or cooling time of
the emitting medium.

The \BBB\ component varies on long time scales, larger than those that
can be investigated by the time span of our observations. It
contributes significantly in the infrared domain, where the amplitude
of its variability is larger than in the optical and ultraviolet
domain. The BATSE light curve also shows a good morphological
resemblance with that of the \BBB\ component.  The only viable
explanation that we have been able to propose for the origin of the
\BBB\ component is the emission from the ``blazar \3c'', a component
that has the same origin as the optical and ultraviolet emission of
3C~279.  This has the consequence that the hard X-ray emission of \3c\ 
observed from the Earth is not the same as the hard X-ray emission
seen by the accretion disk, if reprocessing is to explain the \AAA\ 
component.

While we do not pretend to understand the extremely complex behaviour
of \3c, we can propose a picture that can explain qualitatively the
optical-to-X-ray emission of \3c:
\begin{itemize}
\item The ingredients are a supermassive black hole, a
  geometrically-thick accretion disk, a hard X-ray source located
  inside the funnel, and a jet perpendicular to the disk.
\item A hard X-ray source irradiates the accretion disk, which emits
  the \AAA\ component from the optical domain up to the soft X-ray
  domain. The reprocessing features are considerably attenuated
  because of the reprocessing geometry.
\item If the emission from the accretion disk is removed, one sees the
  ``blazar \3c'', i.e.\ the emission from the jet. This is the \BBB\ 
  component. It dominates the spectral energy distribution of \3c\ at
  long wavelengths (since the far-infrared domain), but also in the
  hard X-ray domain. It may be explained by a SSC model.
\end{itemize}

Seyfert galaxies can be easily linked to the above scenario, if the
\BBB\ component can be suppressed. If they have smaller
accretion-rate-to-Eddington-accretion-rate ratios, their accretion
disks are thinner, which would imply that the curvature detected in
the \AAA\ component would be absent.

\bibliography{/home/astrox/paltani/TeX/biblio}
\bibliographystyle{/home/astrox/paltani/TeX/aa-bib}

\end{document}